\documentclass[conference]{IEEEtran}
\usepackage[LGR,T1]{fontenc}
\usepackage[latin9]{inputenc}
\usepackage{array}
\usepackage{prettyref}
\usepackage{float}
\usepackage{textcomp}
\usepackage{pdfpages}
\usepackage{multirow}
\usepackage{amstext}
\usepackage{graphicx}

\usepackage{listings}
\usepackage{color}

\definecolor{dkgreen}{rgb}{0,0.6,0}
\definecolor{gray}{rgb}{0.5,0.5,0.5}
\definecolor{mauve}{rgb}{0.58,0,0.82}

\begin{document}

\title{DiCE - A Data Encryption Proxy for the Cloud}

\author{\IEEEauthorblockN{Johannes Koppenwallner}
\IEEEauthorblockA{\textit{Faculty of Computer Science} \\
\textit{University of Vienna}\\
Vienna, Austria}
\and
\IEEEauthorblockN{Erich Schikuta}
\IEEEauthorblockA{\textit{Faculty of Computer Science} \\
\textit{University of Vienna}\\
Vienna, Austria \\
erich.schikuta@univie.ac.at}
}

%\author{Johannes Koppenwallner\inst{1} \and Erich Schikuta\inst{1}\orcidID{0000-0002-4126-4243}}

%\authorrunning{J. Koppenwallner et al.}

%\date{\today}

%\institute{University of Vienna, Vienna 1090, Austria\\
%\email{erich.schikuta@univie.ac.at}}

\maketitle

\begin{abstract}
Outsourcing a relational database to the cloud offers several benefits,
including scalability, availability, and cost-effectiveness. However,
there are concerns about the confidentiality and security of the outsourced
data. A general approach here would be to encrypt the data with a
standardized encryption algorithm and then store the data only encrypted
in the cloud. The problem with this approach, however, is that with
encryption, important properties of the data such as sorting, format
or comparability, which are essential for the functioning of database
queries, are lost. One solution to this problem is the use of (e.g. order-preserving) encryption
algorithms, which also preserve these properties in the encrypted
data, thus enabling queries to encrypted data. These algorithms range
from simple algorithms like Caesar encryption to secure algorithms
like mOPE. In order to be able to use these algorithms as easy as
possible, ``DiCE'' a JDBC driver was developed, that parses
SQL queries as a proxy and transparently encrypts and decrypts these
queries. This allows to execute many queries on an encrypted
database in the cloud with (nearly) the performance as on unencrypted
databases. The DiCE driver can be used with any other JDBC driver and
therefore supports a variety of databases. The driver can be configured
to support different encryption algorithms. To keep track of the operations,
the ``Dice Information Client'' has been developed to track the
encryption and decryption of the driver. Although the result of the
performance analysis shows a certain overhead due to the parsing
and encryption of the SQL queries in the Dice driver, this overhead
is significantly reduced by other influencing factors such as the
network and parallel queries.

\end{abstract}


\section{Introduction}

For storing and retrieving structured data, the relational data model
is still the dominant model. More and more data is collected and stored
in databases and they are a critical part in nearly every IT environment.
Traditionally these databases are run in house and managed by members
of the same organization using it. With the rise of cloud computing
this is changing. Databases are outsourced into the cloud and run
and managed by the cloud service provider. This leads
to serious privacy and security concerns, because not only the members
of the organization itself have access to the data, but additionally
the administrators of the cloud provider too. Another serious
concern is that a database, which was formally only accessible in
an internal network, is now accessible over the internet. A solution
to this problem is to encrypt the data with proven secure ciphers
before putting it into the cloud. This approach does not work with
structured data, because important properties of the data are lost
during encryption. The result is that the relational model does not
work for the encrypted data anymore. The format of the data has changed
and queries do not work the way they used to work on the plaintext data.
The data model and any application depending on this schema have to
be changed. Even then, the result comes with a serious performance
penalty by loosing key index properties, which makes this approach often impractical. To avoid this,
other solutions are required. A lot would be gained, if ciphers can
encrypt the values while still keeping format, order or other query
relevant properties. In the optimal case, the data model can be left
unchanged, while still providing data confidentially by encryption.
Of course, any application depending on such an unchanged data model
can be left unchanged too, if encryption and decryption is done transparently.

%\subsection{Goals}
The goal of the DiCE projects aims for
a JDBC proxy which can be used as a drop in for any JDBC driver and
supports different ciphers and encryption models. The solution is
described and all implemented algorithms are shown in detail in this paper. An overview
of the design and implementation is given, and the deployment in the
cloud (AWS) is shown too. Additional ciphers can be added and configured
by a simple user interface. Our approach is evaluated by using
cloud based database as a service model. The performance of specific
queries and general benchmarks (TPC-C~\cite{leutenegger1993modeling}) is evaluated.

The structure of the paper is as follows: In the next section we discuss related
work with a specific focus on similar systems. This is followed by
the presentation of the requirements, design, and implementation of the DiCE system. Our approach is justified and thoroughly evaluated focusing on different aspects of the application domain and infrastructure.
The paper closes by a summary and outlook on future work.

%\subsection{Scope and Limitations}
%
%This paper focuses on relational databases running in the cloud.
%For evaluation the database as a service (DBaaS) model is used from
%some cloud vendors. The encryption of unstructured data or data in
%non-relational databases like NoSQL databases is not examined. Of
%course, some of the presented ciphers may work here as well. The solution
%Dice itself is based on JDBC, which excludes any database not supporting
%it. Encryption is supported for text and integer attributes, other
%data types like date or decimal are not supported. The implemented
%ciphers have these limits regarding the supported data types, too.
%This paper is focused on the encryption of relational databases in
%the cloud, not on security in general. Other important security topics
%like key management, authorization, authentication are only scratched
%on the surface or not discussed at all.

\section{Related Work}\label{relatedwork}

This chapter gives an overview of related work and other existing
solution of encrypted data in the cloud. Although some of these solutions
are applicable for a much broader range of problems, encrypting the
data in relational databases remains here in focus. %For a comprehensive survey on database encryption technologies we refer to~\cite{koppenwallner18}.

\subsection{Eperi Gateway}

Eperi Data Protection for Databases\footnote{http://eperi.de/produkte/database-encryption/}
is an approach to store only encrypted data in the database.
For processing, the data is decrypted by stored procedures so the cleartext
is in memory of DB-Server, thus providing no real end-to-end encryption.
The encryption and decryption is transparent to an application, so
the application does not have to be changed. Eperi gateway
itself provides a lot more features. It is the central point for
accessing and retrieving data from the cloud. It supports transparent
encryption/decryption for different types of data and documents. A disadvantage of this
central gateway approach is, that as all traffic to the cloud comes
and goes through the gateway. Thus, it can be a performance bottleneck and
a single point of failure~\cite{eperi}.
%\begin{figure}[ht]
%\centering
%\includegraphics[width=12cm]{img/website_eperi-gateway-for-databases_schaubild-3}
%
%\caption{Eperi Gateway}
%
%\end{figure}

\subsection{Relational cloud }

Relational cloud\footnote{http://relationalcloud.com} is a project from
MIT to explore and enhance technology of the database as a service
(DBaaS) model in cloud computing. The vision of Relational cloud is
to provide access to all features of a DBMS without the need to manage
hardware, software and privacy.
Relational cloud consists of multiple nodes running a single database
server. Applications communicate by their standard interfaces
like JDBC. A special driver is used to connect to the front-end
to ensure data is kept private. A router is consulted by the front
end to analyze the queries and it determines the execution nodes~\cite{curino2011relational}.

\subsection{CryptDB}

CryptDB~\cite{Popa:2011:CPC:2043556.2043566} is a database management system, which can execute SQL queries
over encrypted data. It follows an SQL-aware encryption strategy and
 evaluates the query directly on the server. The client must only decrypt
the results and does not need to perform any query processing. The encryption can be completely transparent to an application,
as long as the provided client front-end is used. CryptDB uses order
preserving and homomorphic encryption.
%\begin{figure}[ht]
%\centering
%\includegraphics[width=12cm]{img/cryptDB_overview}
%
%\caption{Architecture CryptDB\cite[pp 86]{Popa:2011:CPC:2043556.2043566} }
%\end{figure}

The goal of CryptDB is to be as secure as possible, while still providing
practical access to the database.
However, CryptDB's security was analyzed and some successful attacks were revealed:
\begin{itemize}
\item LP-optimization: Based on combinatorial optimization techniques, this
attack targets deterministic encryption schemes.
\item Sorting attack: This attack decrypts order preserving encrypted columns.
It works on dense sets, where nearly every value exists in the database.
\item Cumulative attack: Another attack on order preserving encrypted columns.
It works even on low-density columns by using combinatorial optimization
techniques.
\end{itemize}

%\subsubsection*{Solutions influenced by CryptDB}

According to~\cite{Tu:2013:PAQ:2535573.2488336}, there exist some
CryptDB based or at least inspired solutions:
\begin{itemize}
\item Monomi~\cite{Tu:2013:PAQ:2535573.2488336}: It is based on the design of CryptDB but with a focus on analytical
queries. It runs queries on encrypted data on top of the PostgreSQL
database. For the execution of complex queries, it uses the split
client/server approach (similar to \cite{Hacigumus:2002:ESO:564691.564717}),
where part of the query is executed on the encrypted data on the server,
and another part of the query is executed on the decrypted data on
the client.
\item Microsoft's Always Encrypted SQL Server~\cite{antonopoulos2020azure}
\item Encrypted Bigquery\footnote{https://github.com/google/encrypted-bigquery-client}: It is a cloud service for analysis of large data-sets from Google storage. For queries it uses an SQL dialect. Encrypted BigQuery is an extension to the Bigquery client, which is capable of client-side encryption for a subset of query types.
\item Skyhigh Networks\footnote{https://www.skyhighsecurity.com}: McAfee Skyhigh Security Cloud is a cloud access security broker (CASB).
This software sits between cloud service consumers and cloud service
providers to enforce security, compliance, and governance policies
for cloud applications. Among tons of other features, it supports
encryption schemes for transparent encrypting of data going through
the broker into the cloud. The Skyhigh Security Cloud supports, besides
regular symmetric encryption schemes, format-preserving and searchable
encryption schemes. Among other schemes, order preserving encryption
is supported too.
\end{itemize}

\subsection{SecureDBaaS}

SecureDBaaS~\cite{jagadeeswaraiah2015securedbaas} differs from other solutions on the application level,
that it does not need a proxy to store metadata. Instead all metadata
is stored encrypted on the server to avoid scalability issues. Features
of SecureDBaaS are
\begin{itemize}
\item guaranteed confidentiality by allowing concurrent SQL over encrypted data,
\item same availability, elasticity and scalability as unencrypted database as a service,
\item concurrent access from distributed clients,
\item no trusted broker or proxy required, and
\item compatible with most relational Databases. It is possible to use existing
database servers like PostgreSQL or MS SQL Server.
\end{itemize}

\subsection{CipherCloud}

CipherCloud\footnote{http://www.ciphercloud.com/} works as gateway which intercepts any traffic
between a database and its clients. In fact, it uses an enhanced JDBC
driver and supports format preserving encryption. It works with Amazon RDS as a database in the cloud~\cite{AWS_data_at_rest}.
%\begin{figure}[ht]
%\centering
%\includegraphics[width=12cm]{img/aws_cipher_cloud_grey}
%
%\caption{Cipher Cloud}
%\end{figure}

\subsection{Voltage Secure}

Voltage Secure\footnote{https://voltage.com} provides another solution for Amazon relational database
service. Unlike CipherCloud it provides its service for applications
in the elastic cloud although with external key management. Similar to CipherCloud,
it is possible with this solution to query over encrypted data~\cite{AWS_data_at_rest}.
%\begin{figure}[ht]
%\centering
%\includegraphics[width=12cm]{img/aws_voltage_grey}
%
%\caption{Voltage Secure}
%\end{figure}

\subsection{Perspecsys}

CloudSOC~\cite{sym_token} is a cloud access security broker (CASB). Part of it (Symantec
Cloud Data Protection \& Security) provides encryption schemes for
the data stored in the cloud. Additionally to the standard schemes,
it supports different functionality (including order) preserving encryption
schemes.

%\subsection{Baseline Research}
Our research in this area is based on our experience with large scale data stores~\cite{schikuta1998vipios} and complex applications in Grids and Clouds~\cite{mach2012generic,schikuta2004n2grid,cs745,weishaeupl2004}, and strongly motivated by our focus on Web-based workflow optimizations~\cite{schikuta2008grid,kofler2009parallel} and their respective management~\cite{stuermer2009building}.

%\begin{figure}[ht]
%\centering
%\includegraphics[width=12cm]{img/mcaffee_comparison_schemes}
%
%\caption{Encryption Strength vs Functionality Preserved (McAfee)}
%\end{figure}
%
%(from The Cloud Encryption Handbook: Encryption Schemes and Their
%Relative Strengths and Weaknesses)
%

%\section{Requirement Analysis}\label{requirementanalysis}

\section{The DiCE System}\label{dice}

This section presents our solution for encrypting
or at least obfuscating data of a database in the cloud. It is called
DICE which is the (clumsy) acronym for ``Database In the Cloud Encrypted''\footnote{The source is available at https://github.com/dicejk/dice}.

\subsection{Requirements}
To achieve the objective of a fully usable encrypted relational database
in the cloud, multiple requirements have to be addressed:

\begin{itemize}
%\item[Cryptography.] Cryptographic techniques and their different characteristics are key for
%the development for a data encryption system for cloud databases.
%For a comprehensive survey on this topic we refer to ???.
%There, a short survey on cryptography is given, which includes history,
%taxonomy and the description of some of the most significant ciphers.
%Ancient ciphers like Caesar's and standard ciphers like DES and AES
%are presented. Attack scenarios and the use of encryption
%to mitigate these threats in the context of a database are described. It shows the
%use case for data at rest and data in transit. For data at rest it
%shows the different levels (storage, database, application) at which
%encryption can be performed. The advantages and disadvantages of the
%place of encryption are discussed here too. Then concrete solutions
%and applications of encryption on the different levels are presented.
%Database specific issues of ciphers are shown, and state of the art
%encryption techniques, such as homomorphic and order preserving encryption
%are described.

\item Relational Model Requirements. The relational model has some implicit requirements which have to
be addressed to be usable. As plain text always satisfies these requirements,
ciphertext does often not (at least not out of the box). As these
requirements are different for each SQL construct, the specific requirements
for data definition and queries are given.

\item Ciphers with Properties.
Although standard ciphers often do not satisfy the properties required
by the relational model, ciphers exist which satisfy some or multiple
of these requirements, as
format preserving, order preserving, functional and homomorphic encryption.
%State of the art ciphers are shown and described in detail.
For queries, order preserving is an often needed property.
Thus, different
order preserving encryption schemes are implemented.
%shown in detail. Security
%definitions are given, and theses ciphers are compared regarding their
%properties, security and implementation.

\item Requirements for Cloud Computing.
Moving data to the cloud requires additional concerns, especially
but not only, regarding security. Contradictions between
security requirements and other cloud-specific requirements, like
scalability or elasticity, have to be addressed. %Deployment scenarios are described,
%and a short overview of available database as a service solutions
%is given. The chapter is concluded with the description of Relational
%Cloud, a project aiming to enhance the existing DBaaS model with security
%and privacy in focus.
\end{itemize}

\subsection{Architecture}

With DiCE the data is always encrypted and decrypted in the application
layer. In the database as a service scenario, this means that the
data is always stored encrypted and never as plaintext in the cloud.
To avoid the effort to change each application for each database at
least for java applications, DiCE is implemented as a JDBC driver.
This makes transparent encryption for the database and the application
possible. As DiCE is in fact a proxy for a real JDBC driver, any application accessing a database via JDBC can use DiCE
to add transparent encryption.

\begin{figure}[htpb]
\centering
\includegraphics[width=6cm]{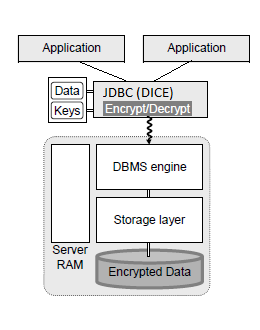}
\caption{DICE Architecture}\label{dicearchitecture}
\end{figure}

An overview of the architecture of DiCE is shown in Figure~\ref{dicearchitecture}.
As shown, DiCE works as layer between the application and the
database. Its main purpose is to transparently encrypt and decrypt
data for an application using it as a JDBC driver. Apart from working
as a JDBC driver, DiCE also provides the functionality for the configuration,
management and traceability of different ciphers and keys.

\subsection{DiCE Components}

DiCE consists of four components, where all its classes are residing
in their respective Java package.
\begin{itemize}
\item dice.sql - This package contains the JDBC proxy driver itself. All necessary
interfaces for JDBC are implemented, and all queries are parsed and
transparently encrypted and decrypted using the ciphers in the package
dice.cipher.
\item dice.cipher - This package contains the ciphers available for dice.
Each cipher provides its own configuration and user interface.
\item dice.info - This package contains everything needed to monitor the execution
of the encryption and decryption in the driver. A small server receives
the messages from the driver. The information can be shown in
client application either using a graphical user interface or the
command line.
\item net.sf.jsqlparser - This external package contains a slightly modified
open source implementation of an SQL parser.
\end{itemize}

\subsection{Supported Ciphers}

DiCE supports different kinds of ciphers with different attributes
regarding order preserving, format preserving and encryption strength.
These are only examples of simple ciphers, but as long as the provided
interface is implemented, any other cipher can be used.
\begin{itemize}
\item Caesar Cipher
\item Generic Substitution Cipher
\item Order Preserving Encryption OPE~\cite{citeulike:9349465} (JOPE was used, a implementation of order-preserving encryption in Java\footnote{https://github.com/ssavvides/jope})
\item Format Preserving Encryption FF1~\cite{Dworkin2013} (FPE4J was used\footnote{https://github.com/zone1511/fpe4j})
\item Advanced Encryption Standard AES~\cite{heron2009advanced} (javax.crypto was used)
\end{itemize}
The first two ciphers are classical ciphers and therefore are not
suitable for serious encryption in the cloud. They are only obfuscating
the data, which can be a wanted feature. However, they are a nice showcase
for interesting properties like order preserving encrypting for data
in a database. OPE~\cite{citeulike:9349465}  is
one of the modern order preserving ciphers. FF1~\cite{Dworkin2013} is
a standard format preserving cipher for numeric values. The last cipher
AES~\cite{heron2009advanced} is one of the standard symmetric block
ciphers, which provides proved strong encryption, but misses important
properties for the general use in a database. If order or format preserving
properties are not needed it is the best choice from an security view
port (ECB mode is possible).

\subsection{Cloud Deployment}

To show the use of DiCE in the cloud, an example configuration of
a database setup using Amazon Web Service is given. Amazon offers
a service called RDS, which stands for ``Relational Database Service''
and is a web service implementing the Database-As-a Service scenario
in the cloud. Different database engines in many configurations are
offered by Amazon, but in this showcase only the smallest configuration using MySQL is chosen.

\section{DiCE User Interface}

DiCE has a minimal user interface developed with Swing for configuration
and for showing the SQL query transformation including encryption
and decryption.

\subsection{Configuration Client}

\begin{figure}[H]
\centering
\includegraphics[width=\columnwidth]{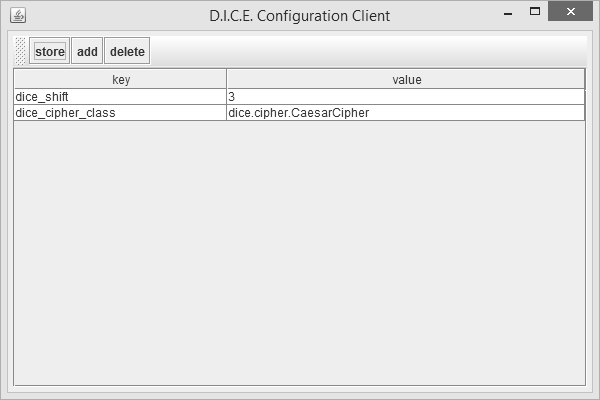}

\caption{DiCE configuration}\label{conf}
\end{figure}

The configuration client is a minimal application to configure DiCE
by simply setting properties for the driver and optionally the used
cipher. As shown in Figure~\ref{conf}  the class of the used cipher
(dice\_cipher\_class) and an parameter for that cipher (dice\_shift)
is set.

\subsection{DiCE Information Client}

\begin{figure}[H]
\centering
\includegraphics[width=\columnwidth]{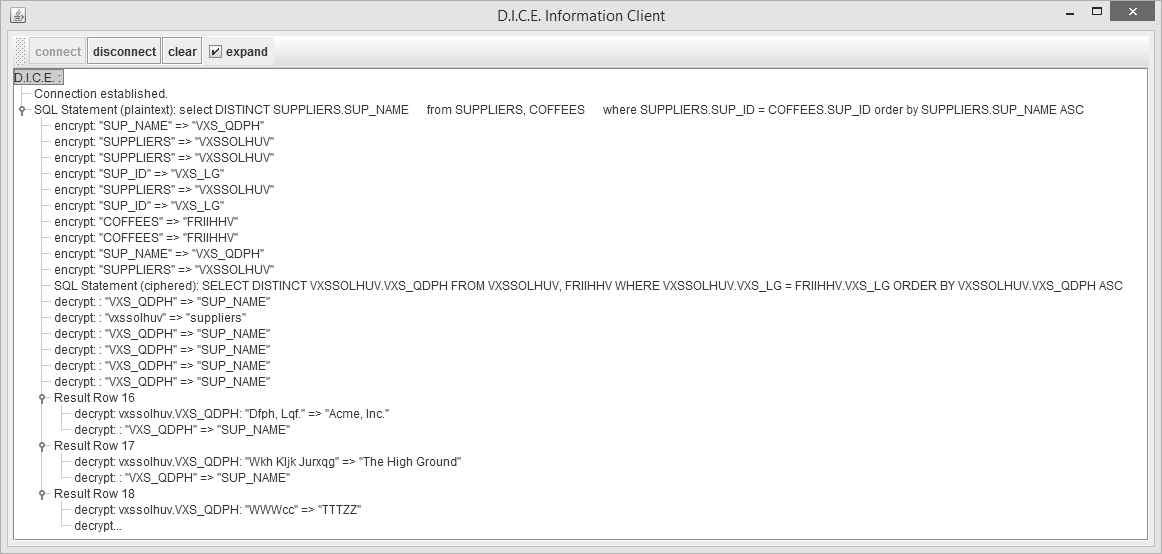}\caption{info client}\label{info}
\end{figure}

Figure~\ref{info} shows the graphical version of the Information Client.
After connecting to the Server, all SQL queries are shown in plain
text and in ciphertext. Additionally all operations of encryption
and decryption are listed.

\subsection{Application Using DiCE As Driver}

\begin{figure}[H]
\centering
\includegraphics[width=\columnwidth]{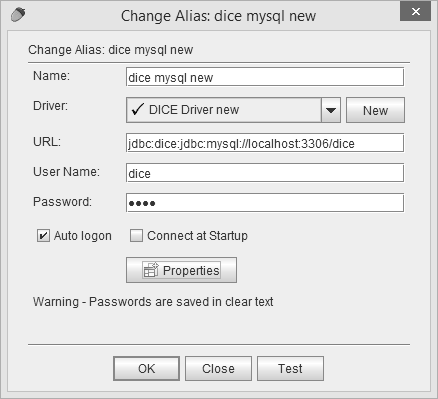}

\caption{Connection with DiCE}\label{conn}
\end{figure}

To show the usage as driver the SQL query tool SQuirreL is used. To
connect to the database the DiCE driver has to be selected and the
URL of the original JDBC driver has to be prefixed by jdbc:dice (as shown in Figure~\ref{conn}). Other settings are the same as for the original driver. SQuirreL SQL\footnote{http://squirrel-sql.sourceforge.net/} was used as SQL client, but
any other supporting JDBC would work too.

\subsection{Performing a query}

\begin{figure}[H]
\centering
\includegraphics[width=\columnwidth]{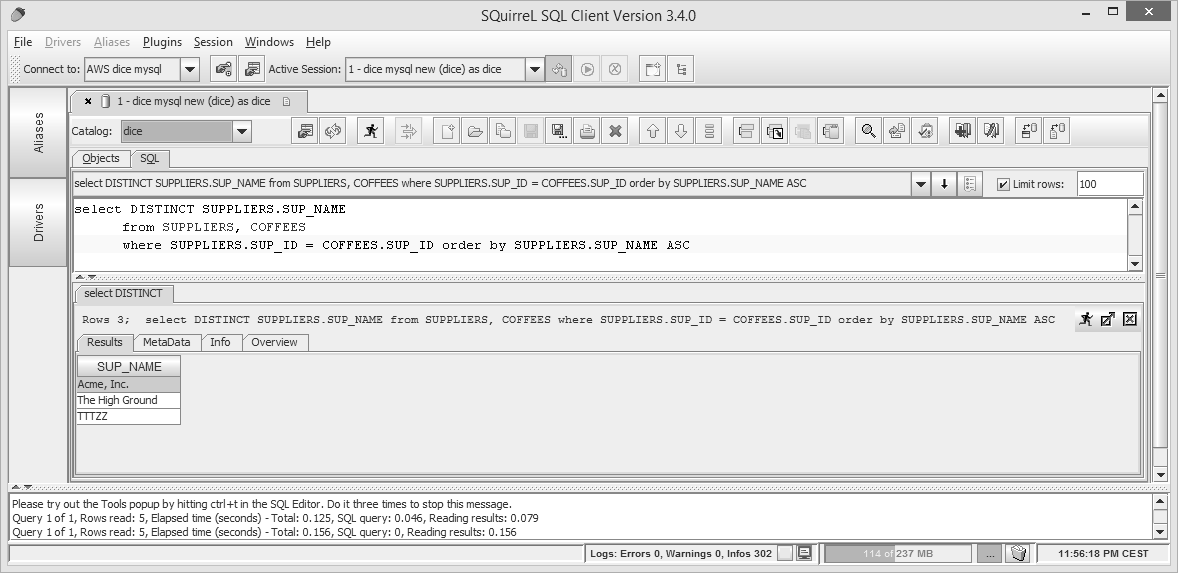}

\caption{SQL Query}\label{query}
\end{figure}

Figure~\ref{query} shows a query using the DiCE driver. As the encryption
and decryption of the query is completely transparent to the application,
no difference between the original driver and the DiCE driver can
be seen.

\subsection{Encrypted Metadata}

\begin{figure}[H]
\centering
\includegraphics[width=\columnwidth]{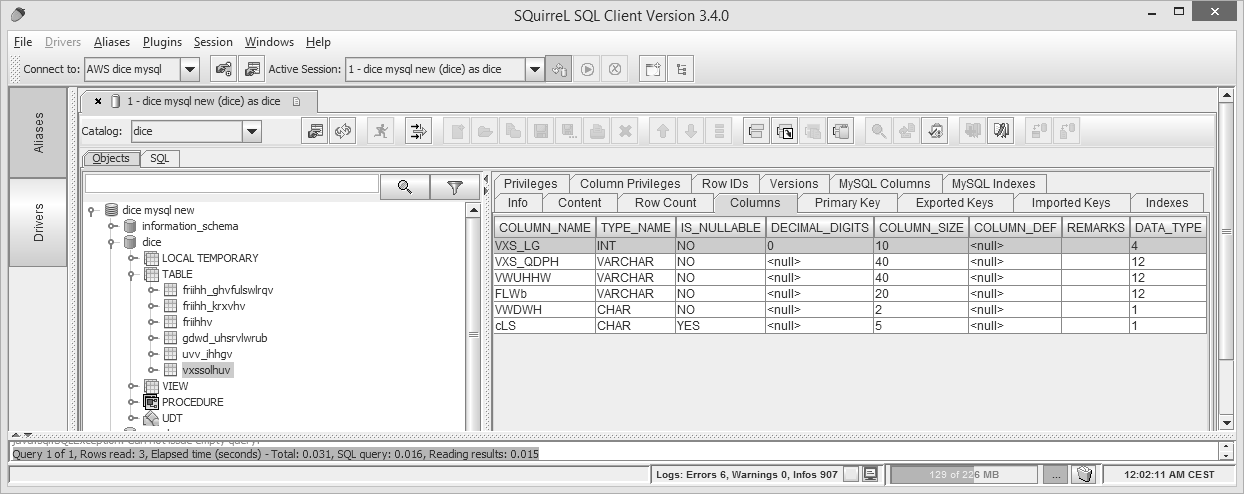}\caption{Schema information}\label{schema}
\end{figure}

Table and column names are encrypted by DiCE too. Figure~\ref{schema}
shows the encrypted metadata in SQuirreL.

\subsection{Encrypted Schema and Result}

\begin{figure}[H]
\centering
\includegraphics[width=\columnwidth]{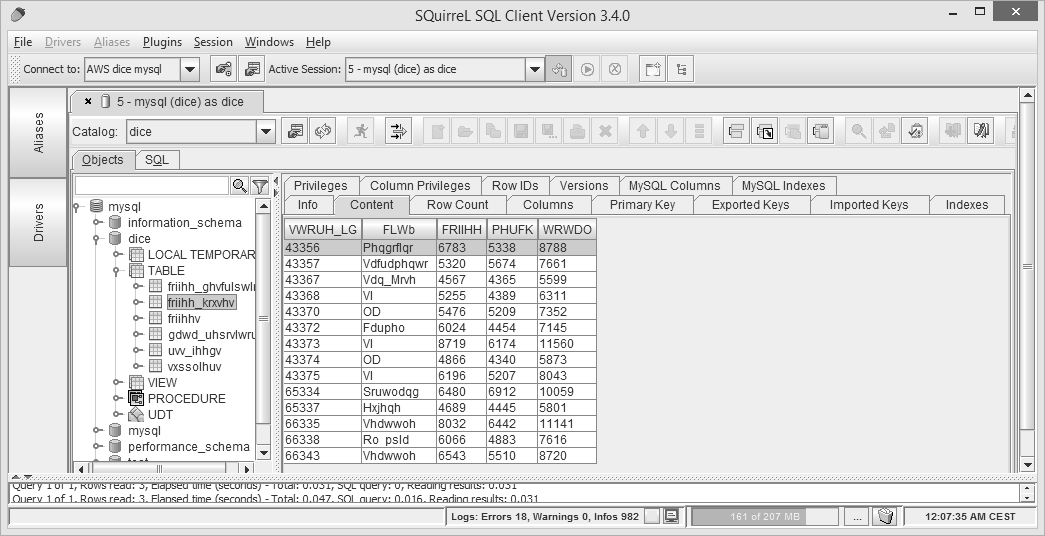}\caption{Content encrypted}\label{encrypted}
\end{figure}

To show the real data and metadata the original driver is used
in the Figure~\ref{encrypted}. This figure shows the encrypted table names and the
content after using DiCE with the Caesar cipher.

\subsection{Decrypted Results}

\begin{figure}[H]
\centering
\includegraphics[width=\columnwidth]{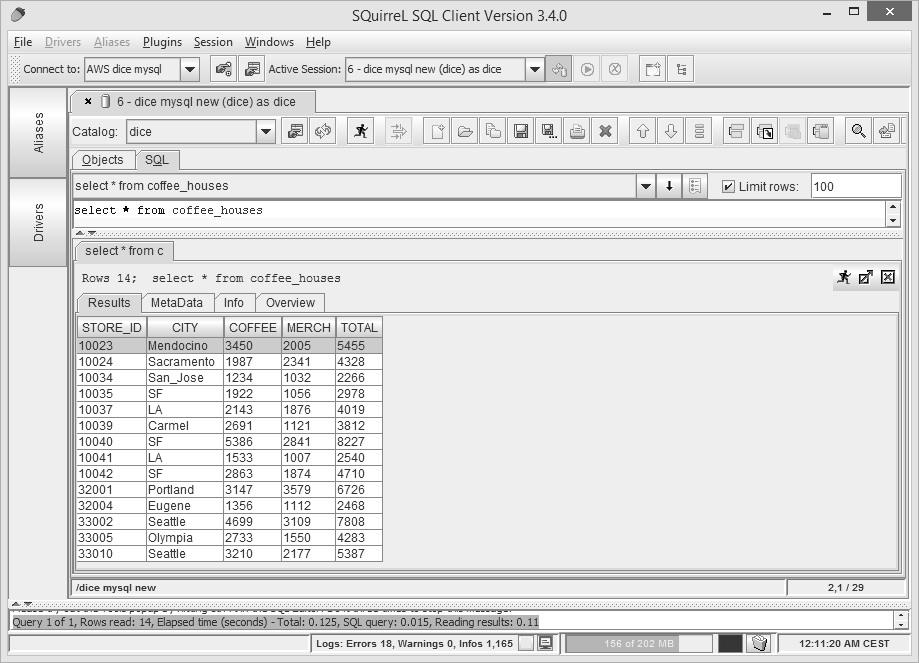}\caption{Content query}\label{decrypted}
\end{figure}

Figure~\ref{decrypted} shows the same result as above, but now using the DiCE
driver, thus transparently decrypting the data.

\section{Evaluation}\label{evaluation}

The performance of DiCE was evaluated in two different
ways. First the standard TPC-C benchmark was used, with an implementation
based on JDBC. %Of course the results are not comparable to the results
%submitted by database vendors, but the results between the same database
%and the use of different JDBC drivers are meaningful.
This benchmark
was executed local (i.e. not in the cloud). Additional to the standard
benchmark the performance of custom analytical queries was analyzed.
A common data model for benchmarks was used and the results were measured
locally and by running the queries on different databases in the cloud.
For the cloud part of the evaluation AWS RDS  was used. As
the encryption and decryption is performed on the client, DiCE was
profiled with the tool VisualVM to show differences between the used ciphers.

\subsection{TPC-C Performance }

The TPC-C benchmark simulates a complete business cycle. Multiple
business transaction are executed beginning with creation of orders,
querying of existing orders, payment by customers and observing stock
levels of warehouses. The metric for the TPC-C benchmark is the ``Maximum
Qualified Throughput'' (MQTh). It is the number of orders processed
per minute.

%\subsubsection{TCJ}

TCJ (TPC-C via JDBC) is an implementation of the TPC-C benchmark in
Java. It uses JDBC to connect to the database\footnote{http://sourceforge.net/projects/tpcc-jdbc/}.
The benchmark was executed with the following parameters:
\begin{itemize}
\item -w 10 (number of warehouses)
\item -r 1 (length of ramp-up phase)
\item -m 10 (length of measurement interval)
\end{itemize}
It was executed with the plain JDBC driver and with the DiCE driver
using a simple rotating cipher. It is interesting to see that although
the results (the throughput) are very similar, the curves look differently.
Using the DiCE driver results in a 12\% longer run time. High throughput
is reached after 120 seconds, whereas the standard driver reaches
the high throughput already after 60s. It shows that the overhead
of the SQL parsing in the DiCE driver has impact on the response time
of every single transaction, but only a small impact on the throughput.

\subsubsection{Throughput using the MySQL driver}

\begin{figure}[H]
\centering
\includegraphics[width=\columnwidth]{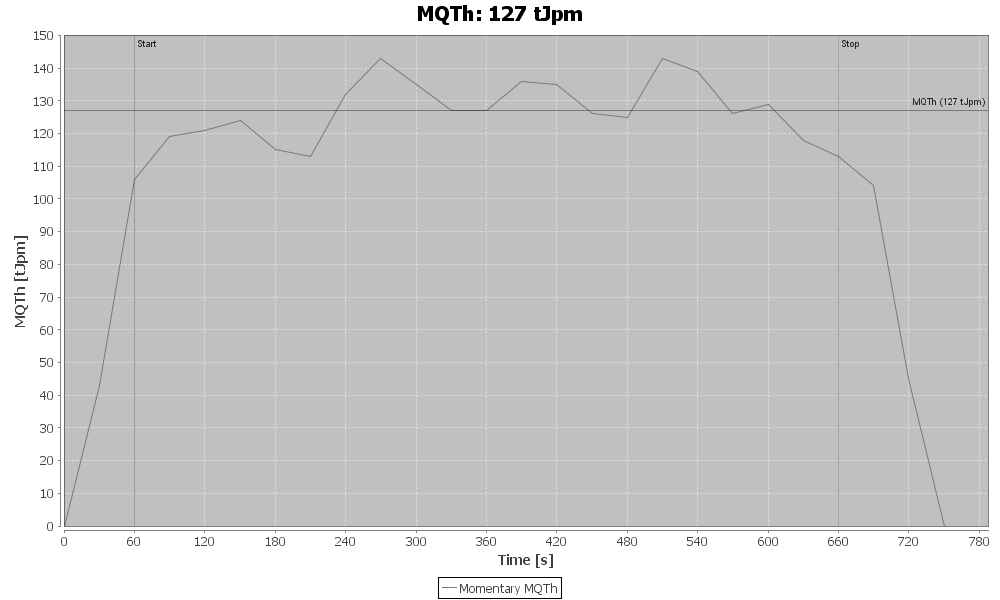}\caption{TCJ plain MySQL driver}\label{tcjplain}
\end{figure}

Using the MySQL JDBC driver a throughput of 127 MQTh was measured.
The run time took 750 seconds and the peak throughput was 143 MQTh (see Figure~\ref{tcjplain}).

\subsubsection{Throughput using the DiCE driver}

\begin{figure}[H]
\centering
\includegraphics[width=\columnwidth]{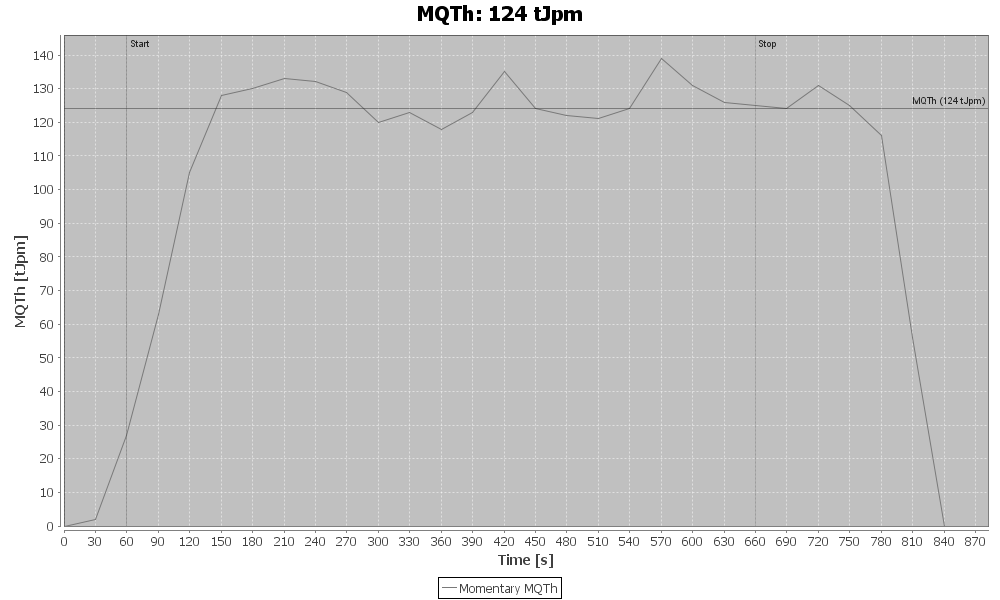}

\caption{TCJ DiCE using rotating cipher}\label{tcjdice}
\end{figure}

Using the DiCE driver with the rotating cipher a throughput of 124
MQTh was measured. The run time took 840 seconds and the peak throughput
was 143 MQTh (see Figure~\ref{tcjdice}).

\subsection{Query Evaluation }

\begin{figure}[H]
\centering
\includegraphics[width=\columnwidth]{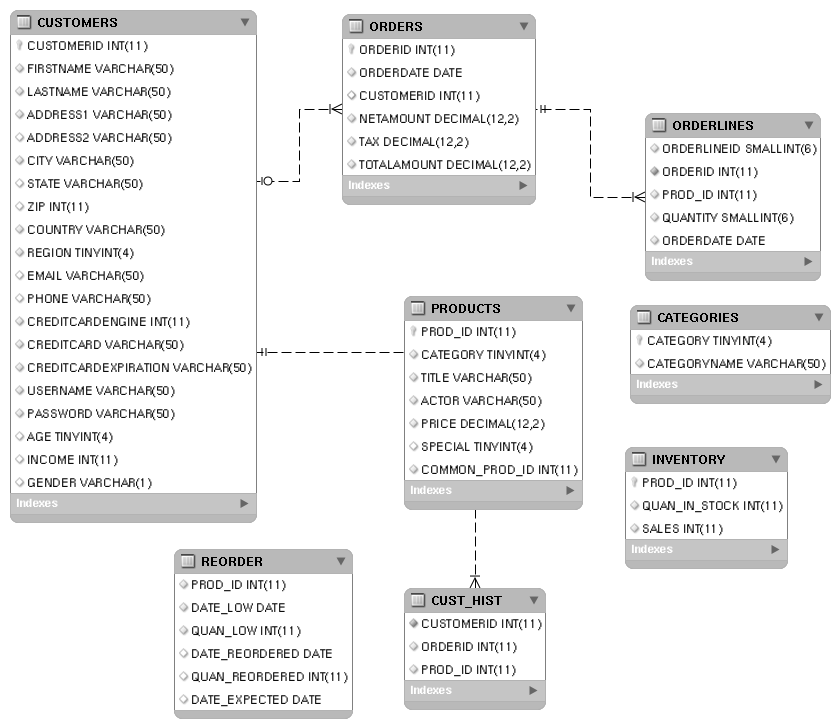}

\caption{DVD Store ER-model}\label{dvder}
\end{figure}

As sample data model the Dell DVD Store\footnote{https://linux.dell.com/dvdstore/}
was used. The Entity-Relationship diagram is shown in Figure~\ref{dvder}.

\subsubsection{Selected SQL Queries}

As each cipher determines the properties of the ciphertext, we used
a selection of SQL queries to show the operations
supported on the ciphertext. For example, if a query on
the plain text requires the operator ``>'', this operator has to be supported
on the ciphertext as well to retrieve the correct result. The SQL queries comprised:

\begin{itemize}
\item{Simple Query with Projection}

\begin{lstlisting}
SELECT C.FIRSTNAME, C.LASTNAME
FROM CUSTOMERS C
WHERE C.CUSTOMERID = '123456789'
\end{lstlisting}

\item{Query with Join}

\begin{lstlisting}
SELECT DISTINCT C.FIRSTNAME, C.LASTNAME
FROM CUSTOMERS C INNER JOIN ORDERS O ON C.CUSTOMERID = O.CUSTOMERID
WHERE O.ORDERDATE > '2010-12-30'
\end{lstlisting}

\item{Query with Order by }

\begin{lstlisting}
SELECT FIRSTNAME,LASTNAME,AGE
FROM CUSTOMERS C
ORDER BY C.AGE LIMIT 5
\end{lstlisting}

\item{Query with Where clause =, >,< }

\begin{lstlisting}
SELECT *
FROM CUSTOMERS C
WHERE C.INCOME < 20001
\end{lstlisting}

\item{Query with Between }

\begin{lstlisting}
SELECT count(*)
FROM CUSTOMERS C
WHERE C.INCOME BETWEEN 30000 AND 40000
\end{lstlisting}

\item{Query with Like }

\begin{lstlisting}
SELECT *
FROM CUSTOMERS C
WHERE C.LASTNAME LIKE '%MEI%'
\end{lstlisting}

\item{Query with Aggregate (avg,max,sum) }

\begin{lstlisting}
SELECT MAX(C.INCOME)
FROM CUSTOMERS C
\end{lstlisting}

\item{Query with Functions}

\begin{lstlisting}
SELECT *
FROM CUSTOMERS C
WHERE C.INCOME = C.AGE * 1000
\end{lstlisting}

\item{Query with Group by, having}

\begin{lstlisting}
SELECT SUM(O.QUANITY)
FROM ORDERLINES O
GROUP BY PROD_ID HAVING ORDERDATE > '15-12-2017'
\end{lstlisting}
\end{itemize}

\subsubsection{Local Evaluation}

The evaluation has been performed on a Lenovo Thinkpad T460s running
windows 10. The relational database for local queries has been installed
on a virtual machine running Linux. All evaluation programs and the
DiCE driver itself have been running under Windows directly avoiding
the bias of a virtual environment.

\paragraph{Database Load}

\begin{figure}[H]
\centering
\includegraphics[width=\columnwidth]{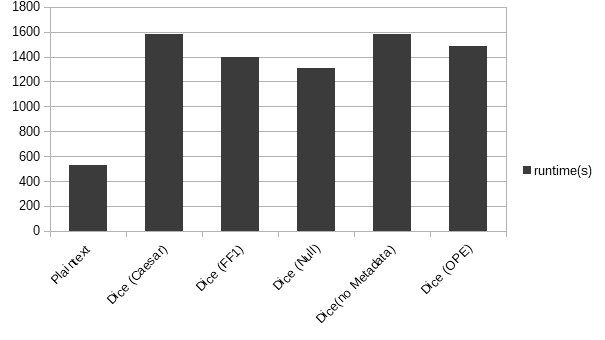}

\caption{\label{fig:Import-run-time}Import run time }

\end{figure}

The first test was simple. We inserted the data into the local database.
A small database with the size of 100MB was created. The
evaluation was executed with the standard JDBC driver (plain text)
and the DICE driver with various ciphers. The run time with DiCE is
about 2,5x times longer with the use of a dummy cipher and about 3x
longer with a real cipher. The overhead of the ciphers is small compared
to the overhead of parsing and rewriting the SQL statement (see Figure~\ref{fig:Import-run-time}).

\begin{figure}[H]
\centering
\includegraphics[width=\columnwidth]{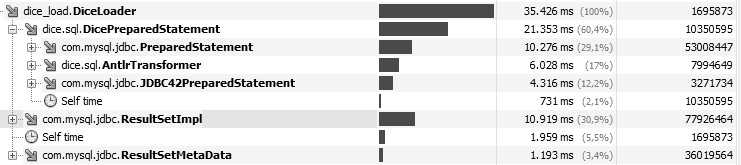}

\caption{Profiler Load DiCE Caesar}\label{load}
\end{figure}

\begin{figure}[H]
\centering
\includegraphics[width=\columnwidth]{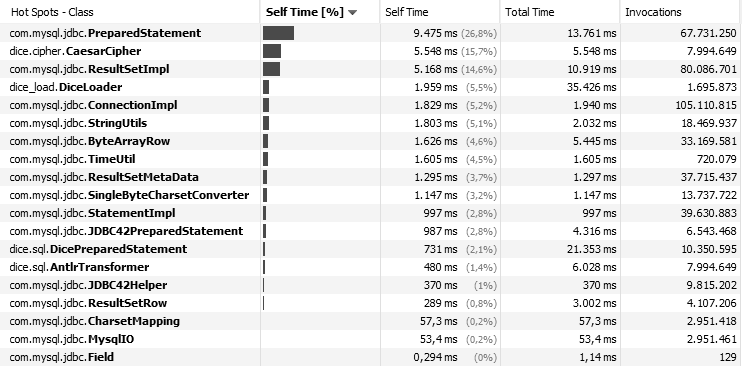}

\caption{Profiler Hotspots Load Caesar}\label{hotspots}

\end{figure}

Using local networking and VisualVM 1.3.9 as profiler for DiCE with
Caesar cipher, the following results were retrieved: It shows the
significant overhead of DiCE. The PreparedStatement itself is about
1/3 of the run time. The rest is the DiCEPreparedStatement where the
cipher itself costs about 16\% of the total run time (see Figures~\ref{load} and \ref{hotspots} respectively).

\paragraph{SQL Queries}

\begin{figure}[H]
\centering
\includegraphics[width=\columnwidth]{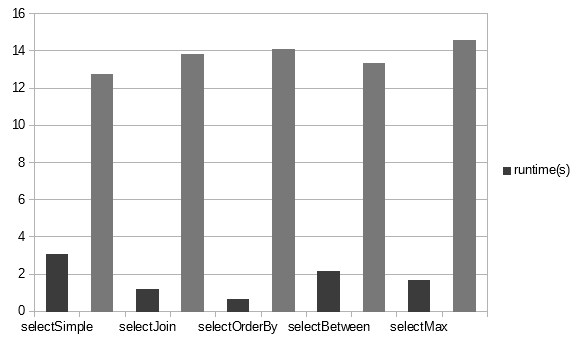}

\caption{Queries local DiCE vs plain text}\label{queries}
\end{figure}

The queries were evaluated comparing plain
text and DiCE. The result shows a significant overhead of DiCE. It is interesting to see, that while the queries without DiCE differ
large in run time (0,71s - 3,128s), the relative difference of the
run time of the queries with DiCE is much smaller, because the parsing
overhead and additional processing overhead of the JDBC proxy is nearly
constant (see Figure~\ref{queries}).

The absolute numbers are shown in Tables~\ref{dicecaesar} and  \ref{plainmysql} respectively.

\begin{table}[H]

\caption{DiCE Caesar}\label{dicecaesar}

\centering
\begin{tabular}{|c|c|c|c|}
\hline
Query & Start & End & Diff (sec)\tabularnewline
\hline
\hline
selectSimple & 1493578154043 & 1493578166816 & 12,773\tabularnewline
\hline
selectJoin  & 1493578293433 & 1493578307320 & 13,887\tabularnewline
\hline
selectOrderBy & 1493578418780 & 1493578432914 & 14,134\tabularnewline
\hline
selectBetween & 1493578504800 & 1493578518261 & 13,461\tabularnewline
\hline
selectMax  & 1493578555254 & 1493578569857 & 14,603\tabularnewline
\hline
\end{tabular}
\end{table}

\begin{table}[H]
\caption{Plain MySQL}\label{plainmysql}
\centering
\begin{tabular}{|c|c|c|c|}
\hline
Query & Start & End & Diff (sec)\tabularnewline
\hline
\hline
selectSimple & 1493578737742 & 1493578740870 & 3,128\tabularnewline
\hline
selectJoin  & 1493578818546 & 1493578819744 & 1,198\tabularnewline
\hline
selectOrderBy & 1493579092716 & 1493579093426 & 0,710\tabularnewline
\hline
selectBetween & 1493578985059 & 1493578987229 & 2,170\tabularnewline
\hline
selectMax  & 1493578932817 & 1493578934526 & 1,709\tabularnewline
\hline
\end{tabular}

\end{table}

\paragraph{Naive Implementation vs DICE Caesar}

\begin{figure}[H]
\centering
\includegraphics[width=\columnwidth]{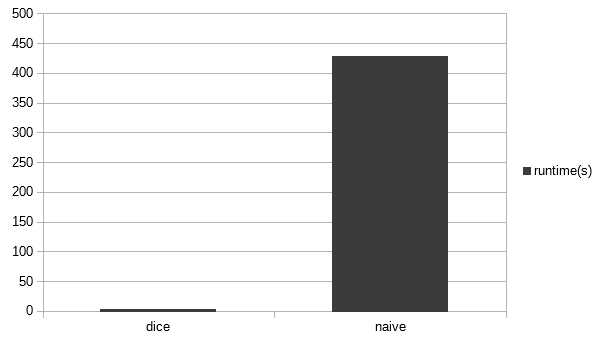}

\caption{Run time DiCE vs Naive}\label{dicevsnaive}
\end{figure}

The last local evaluation was to run the queries with DiCE and compare
them against a naive implementation, where no properties in the encrypted
database are preserved. Even for a small data set the naive implementation
is 2 orders of magnitude slower than the order preserving implementation (see Figure~\ref{dicevsnaive}).

The absolute numbers are shown in Tables~\ref{tdicevsnaive}.

\begin{table}[H]
\caption{DiCE vs Naive}\label{tdicevsnaive}
\centering
\begin{tabular}{|c|c|c|c|}
\hline
Implementation & Start & End & Diff(sec)\tabularnewline
\hline
\hline
naive & 1493672968762 & 1493673398239 & 429,477\tabularnewline
\hline
order preserving & 1493673945301 & 1493673950143 & 4,842\tabularnewline
\hline
\end{tabular}

\end{table}

\subsubsection{Multiple local concurrent connections}

\begin{figure}[H]
\centering
\includegraphics[width=\columnwidth]{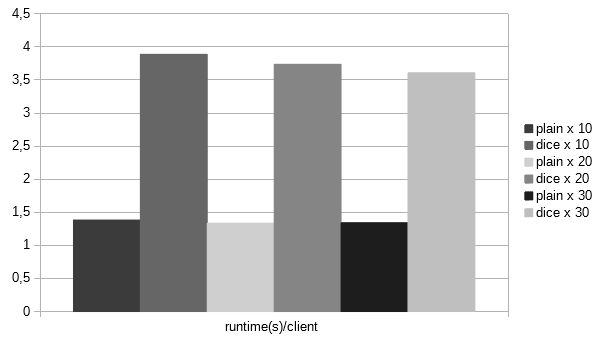}

\caption{multiple local connections}\label{multiplelocalconnections}
\end{figure}

The last local evaluation shows the impact of using multiple local
clients accessing a local database. The run time per client stays
constant and does not scale (or only slightly). It seems like 10 clients
and the server are enough to utilize the machine to capacity (see Figure~\ref{multiplelocalconnections}).

\subsection{Cloud Evaluation}

\subsubsection{Cloud Service Provider}

The evaluation in the cloud was performed using the AWS relational
database service. The same select-statements were performed on the plain text
and the encrypted data. Additionally to that, ``naive queries''
were performed for comparison. These are different queries, because
they represent the case where no properties are preserved by the encryption.
Thus, the whole data as to be retrieved and decrypted before processing
the result. To import the data of the local databases in the cloud
databases \texttt{mysqldump} was used.
%For example \texttt{mysqldump -u dice -{}-databases
%DICE\_C -{}-single-transaction -{}-compress -{}-order-by-primary -p<password>
%| mysql -u dice -{}-port 3306 -{}-host=dice.cra9kiqswfw9.eu-central-1.rds.amazonaws.com
%-p<password>} exported the local database DICE\_C to the cloud database
%DiCE.
This is much faster than to insert the data with JDBC again.

\subsubsection{Plain and DiCE vs Naive}

\begin{figure}[H]
\centering
\includegraphics[width=\columnwidth]{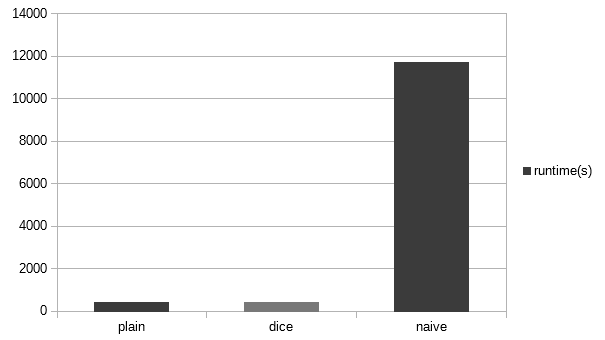}

\caption{Evaluation Cloud DiCE vs Plain}\label{evaluationclouddicevsplain}
\end{figure}

For the first tests the smallest configuration AWS db.t2.micro was
used. It features 1 CPU, 1 GB RAM, SSD as storage and was running in
region EU (Frankfurt). MySQL 5.6.27 was used as database. All tests
where performed multiple times, to get more accurate results. The performance is depicted in Figure~\ref{evaluationclouddicevsplain}.

It is interesting to see that there is nearly no difference between
DiCE and the plain text anymore. The naive approach is clearly inferior,
with a run time about 26x slower than the others. CPU utilization
of the database was about 60\%.

The absolute numbers are shown in Tables~\ref{cloudplaindata}, \ref{clouddicedata}, and \ref{cloudnaivedata} respectively.

\begin{table}[H]
\caption{Cloud plain data}\label{cloudplaindata}
\centering
\begin{tabular}{|c|c|c|c|}
\hline
plain & start & end & run time\tabularnewline
\hline
\hline
 & 1494174290170 & 1494174758243 & 468,073\tabularnewline
\hline
 & 1494174828367 & 1494175303735 & 475,368\tabularnewline
\hline
 & 1494180100245 & 1494180588455 & 488,210\tabularnewline
\hline
 & 1494180669543 & 1494181148966 & 479,423\tabularnewline
\hline
\end{tabular}
\end{table}

\begin{table}[H]
\caption{Cloud DiCE}\label{clouddicedata}
\centering
\begin{tabular}{|c|c|c|c|}
\hline
dice & start & end & run time\tabularnewline
\hline
\hline
 & 1494175776017 & 1494176254981 & 478,964\tabularnewline
\hline
 & 1494176663187 & 1494177148177 & 484,990\tabularnewline
\hline
 & 1494177277103 & 1494177744003 & 466,900\tabularnewline
\hline
 & 1494179453568 & 1494179907970 & 454,402\tabularnewline
\hline
\end{tabular}
\end{table}

\begin{table}[H]
\caption{Cloud naive}\label{cloudnaivedata}
\centering
\begin{tabular}{|c|c|c|c|}
\hline
naive & start & end & run time\tabularnewline
\hline
\hline
 & 1494174471845 & 1494186248363 & 11776,518\tabularnewline
\hline
 & 1495559376409 & 1495571426342 & 12049,933\tabularnewline
\hline
\end{tabular}
\end{table}

\subsubsection{Multiple parallel Clients}

To get more load on the server we used the same configuration as before. OPE as cipher was used, but this time with multiple clients (all running on the same machine) querying the server simultaneously.
This resulted in higher load on the server (80\% and more) and showed
the same effect as in the TPC-C benchmark: the average run time per
client decreased. The load on the client and the network was negligible (see Figure~\ref{ope1}).

\begin{table}[H]
\centering
\includegraphics[width=\columnwidth]{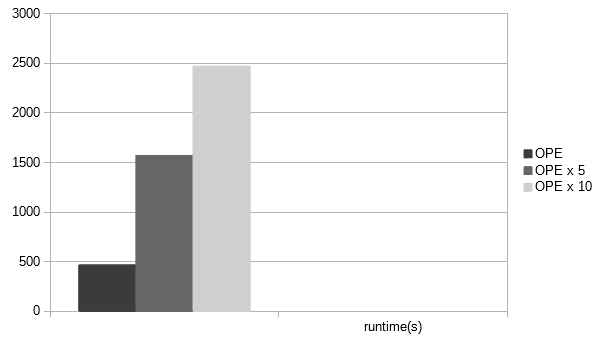}
\caption{OPE multiple clients}\label{ope1}
\end{table}

Although the total run time is longer, the run time per client is
decreasing (see Figure~\ref{ope2}):

\begin{table}[H]
\centering
\includegraphics[width=\columnwidth]{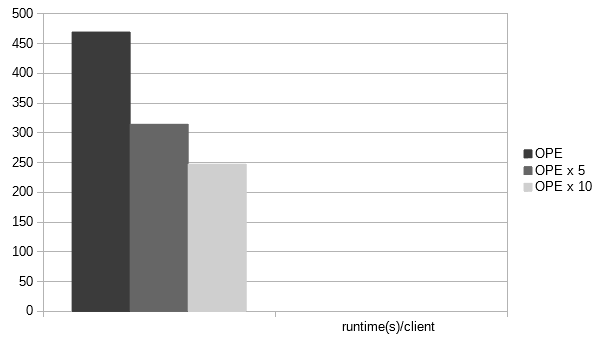}

\caption{OPE per client}\label{ope2}
\end{table}

\subsubsection{Different Cloud Configurations}

\begin{figure}[H]
\centering
\includegraphics[width=\columnwidth]{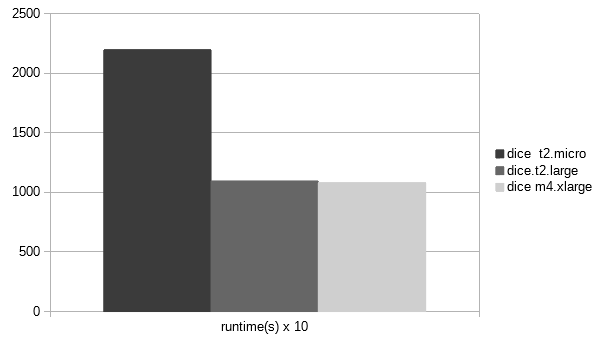}

\caption{DiCE cloud scale}\label{cloudscale}
\centering
\end{figure}

To examine the impact of different RDS configurations 3 configurations were selected:
\begin{itemize}
\item t2.micro (1 CPU, 1GB RAM)
\item t2.large (2 CPU, 8GB RAM)
\item m4.large (4 CPU, 16GB RAM)
\end{itemize}
To avoid the network as bottleneck, these tests were executed with
a 100 MBiT network connection to the internet. Whereas the more powerful
configurations perform the same, the run time of the clients querying
the smallest database is halved (see Figure~\ref{cloudscale}).

\begin{figure}[H]
\centering
\includegraphics[width=\columnwidth]{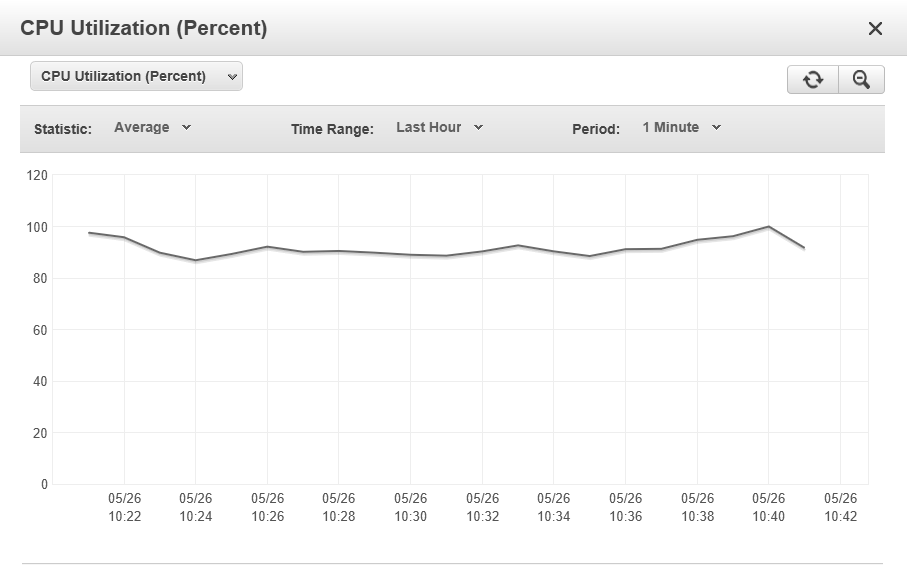}

\caption{DICE\_C T1 micro}\label{t1micro}
\end{figure}

To confirm the assumption, the CPU utilization was monitored, showing
nearly 100\% CPU workload, as shown in Figure~\ref{t1micro}.

\subsection{Findings}\label{findings}

First of all, the naive approach of selecting all data, decrypting
it on the client and filtering it in the program, does simply not
work. Even for a tiny data model, these queries take between 26x and
88x longer than by using the order preserving encryption of DiCE.
Thus, this is a no viable solution for using an encrypted database in the
cloud. However, the results of DiCE are different.

\subsubsection{DiCE Performance}

DiCE has a nearly constant overhead for parsing each query and passing
it to the real JDBC driver. This overhead is significant and has a
great impact on the response time of an individual query. The impact
is between 4x - 18x depending on the query. This does not sound good,
but if multiple queries are executed in parallel, this overhead is
reduced to about 2.5x in total. The impact
on the throughput measured with TCJ is even smaller. The throughput
here is only 3\% less than without encryption, while the overall run-time
is about 15\% longer. This is consistent with the results of the selected
queries, which show that the impact on throughput is much smaller
than on response time. But beside the parsing and processing of the
SQL in the JDBC proxy, the following factors also have significant
impact on the overall performance.
\begin{itemize}
\item Used Cipher
\item Encrypted Metadata
\item Local / Network
\end{itemize}

\paragraph{Impact of used cipher}

The ciphers had only a small impact on the overall performance. The
difference between them was only about 15\%. Only the dummy implementation
of a cipher (NullCipher), was about 30\% faster than the slowest real
cipher. This can be explained by the big influence of the parsing
on performance, which reduces the impact of the chosen cipher on the
overall performance. These results are only valid for the tested ciphers.
Use of a more secure but less efficient cipher can dramatically
change these results. \label{par:Impact-of-the}

\paragraph{Impact of metadata encryption}

The encryption or not encryption of metadata, as encrypting table
or attribute names, had no measurable impact on the performance. The
reason for this is that metadata is only retrieved once per table,
which causes only very few calls compared to calls for encrypting
and decrypting the normal data.

\paragraph{Impact of the network (Cloud)}

The network has a big impact on the overall performance. The penalty
is highly dependent on the speed of the network and by the amount
of data transferred between the client and the server in the cloud.
In the tests performed, the influence on the overall performance was
so high, that any run-time differences between DICE and the direct
use of the native driver nearly vanished complete.

\subsubsection{Using DiCE in the Cloud}

The performance results for using an encrypted database with DiCE
in the cloud look promising. Yes, there is an overhead, but it is
almost negligible compared to deploying the data without encryption
in the cloud.

\section{Future Work and Outlook}

Currently, the supported ciphers for DiCE are AES as an example for
a cipher without properties, JOPE as an example of order preserving
encryption and FF1 as example for format preserving encryption. Additionally
classic algorithms like Caesar's and generic rotation ciphers exist.
This leaves room for more ciphers for test. As of the time of writing,
the implementation is limited to string and number types, so the support
of other types, like date or decimal, would be interesting. JDBC is
supported, but the other common database interface ODBC is not supported.
Not all types and SQL statements are supported, it would be interesting
in supporting for example LIKE (which requires a searchable encryption
scheme) or aggregate functions like SUM (which requires a homomorphic
encryption scheme). The user interface is currently the absolute minimum
to be usable. In fact it just sets the necessary system properties.
Here, a lot of improvements are possible. Another open question touches
security. Is there any evidence that in house administrated databases
are more secure than outsourced databases? Is the database a service
scenario eventually even more secure than the in house alternatives.

%\subsection{Outlook}

Optimal order preserving encryption is here, but it has to prove itself
over the years. As of today, no high quality standard implementation, as for example for AES, exists. This will hopefully change soon.

The
big elephant in the room is fully homomorphic encryption, and future
will show, if it is possible to develop a secure and well performing
cipher. If it is possible to achieve this, the impact goes far beyond
database encryption, because then it would be possible to directly
operate on the encrypted data and move the whole data processing in
the cloud, too.

Quantum computing could be a disruptive technology
for many ciphers, which are currently believed to be secure.

The adoption
of cloud computing is continuing to grow, but as more and more important
services rely on it, it seems very likely that more legislative control
will be applied to it. Privacy and security in and outside the cloud
will be more important than ever, and legislative regulation like
GDPR will become more and more relevant.

\section{Conclusion}\label{conclusion}

Collecting and storing only the minimal required data seems obvious,
but often it is not. The GDPR (General Data Protection Regulation)~\cite{hoofnagle2019european}
makes data minimization of personal
data a principle and from a security point of view this makes absolute
sense. Never collected data does not have to be stored, maintained
and protected. But there are enough systems, where sensible data is
definitely needed. Here it is crucial to make this classification
as sensible data explicit. Without this knowledge, it is impossible
to protect it accordingly. Mixing sensitive with non sensitive data
is also a bad idea, because then everything has to be handled as sensitive
data and this comes with an overhead. Often it is good practice to
separate sensitive tables or even systems from non sensitives, but
this will work only if considered from start.

This paper discussed
many ciphers, with different properties and different strength of
security. It is recommended to always go for the best matching cipher
regarding security requirements and needed functionality, even if
that means to use multiple different ciphers. Here, the use of the
smallest common denominator is definitely no good idea, as there is
no perfect cipher which is best for all use cases. Ciphers with specific
properties required for queries on encrypted data are available and
usable. Order preserving encryption schemes with best possible security
exist, but they are still not as secure as algorithms without additional
properties. It is important to know, that these ciphers are relatively
new and not analyzed in depth like standard algorithms as AES, for
example. Of course often not optimal encryption is better than none,
but it is important to know about the weaknesses of these ciphers. As a
result, is not recommended to use any non standard encryption schemes
for high classified data.

The use of these ciphers also does not come
for free. There is a certain overhead by embedding the encryption
in a JDBC driver compared to using the native JDBC driver of the database
directly. The reason for this is that much more processing has to
be done before sending the query to the server. The SQL statement
has to be parsed, encrypted and sometimes rewritten before it is forwarded
to the real driver. The results of a query have to be parsed and decrypted
again. The performance results show that the response time is significant
higher, but by simulating multiple simultaneous queries the impact
on the throughput is not as dramatic. Compared with the naive approach,
by not enabling joins or indices on the encrypted data the result
is clear: The naive approach is simple not working for anything but
toy projects.

As DiCE makes it possible to use any database supporting
JDBC, it was good to see that MySQL and Microsoft SQLServer worked
as expected. Of course, as DiCE uses its own SQL parser it is necessary
to use standard SQL, even if the underlying driver would support some
specific extensions of the standard. Making the ciphers interchangeable
makes it easy to implement and evaluate new ciphers. So DiCE can be
also seen as a test bed for some property preserving ciphers. Making
this more adaptable for specific columns and multiple ciphers simultaneously
would be a good feature, too.

Of course, it can not repeated more often
that security is more than encryption, and a lot more than encrypting
the database has to be done to achieve security. Using the database
as service solutions from AWS and Azure worked like a charm. It literally
takes only 5 minutes and a database is up and running and available
for operation.

Summing up, DICE proved as viable solution for encrypting database data at rest in the Cloud and the additional costs are acceptable.

%Network access and speed is definitely an issue. For
%example the data setup for evaluation took some time to get it into
%the cloud.

%\bibliographystyle{IEEEtran}
%\bibliography{dice}

\begin{thebibliography}{10}
\providecommand{\url}[1]{#1}
\csname url@samestyle\endcsname
\providecommand{\newblock}{\relax}
\providecommand{\bibinfo}[2]{#2}
\providecommand{\BIBentrySTDinterwordspacing}{\spaceskip=0pt\relax}
\providecommand{\BIBentryALTinterwordstretchfactor}{4}
\providecommand{\BIBentryALTinterwordspacing}{\spaceskip=\fontdimen2\font plus
\BIBentryALTinterwordstretchfactor\fontdimen3\font minus
  \fontdimen4\font\relax}
\providecommand{\BIBforeignlanguage}[2]{{%
\expandafter\ifx\csname l@#1\endcsname\relax
\typeout{** WARNING: IEEEtran.bst: No hyphenation pattern has been}%
\typeout{** loaded for the language `#1'. Using the pattern for}%
\typeout{** the default language instead.}%
\else
\language=\csname l@#1\endcsname
\fi
#2}}
\providecommand{\BIBdecl}{\relax}
\BIBdecl

\bibitem{leutenegger1993modeling}
S.~T. Leutenegger and D.~Dias, ``A modeling study of the tpc-c benchmark,''
  \emph{ACM Sigmod Record}, vol.~22, no.~2, pp. 22--31, 1993.

\bibitem{eperi}
Eperi, ``How eperi can help to achieve compliance with strict data residency
  requirements,'' [Online]. Available: https://blog.eperi.co, 2018.

\bibitem{curino2011relational}
C.~Curino, E.~Jones, R.~A. Popa, N.~Malviya, E.~Wu, S.~Madden, H.~Balakrishnan,
  and N.~Zeldovich, ``{Relational Cloud: A Database Service for the Cloud},''
  in \emph{5th Biennial Conference on Innovative Data Systems Research},
  Asilomar, CA, January 2011.

\bibitem{Popa:2011:CPC:2043556.2043566}
\BIBentryALTinterwordspacing
R.~A. Popa, C.~M.~S. Redfield, N.~Zeldovich, and H.~Balakrishnan, ``Cryptdb:
  Protecting confidentiality with encrypted query processing,'' in
  \emph{Proceedings of the Twenty-Third ACM Symposium on Operating Systems
  Principles}, ser. SOSP '11.\hskip 1em plus 0.5em minus 0.4em\relax New York,
  NY, USA: ACM, 2011, pp. 85--100. [Online]. Available:
  \url{http://doi.acm.org/10.1145/2043556.2043566}
\BIBentrySTDinterwordspacing

\bibitem{Tu:2013:PAQ:2535573.2488336}
\BIBentryALTinterwordspacing
S.~Tu, M.~F. Kaashoek, S.~Madden, and N.~Zeldovich, ``Processing analytical
  queries over encrypted data,'' \emph{Proc. VLDB Endow.}, vol.~6, no.~5, pp.
  289--300, Mar. 2013. [Online]. Available:
  \url{http://dx.doi.org/10.14778/2535573.2488336}
\BIBentrySTDinterwordspacing

\bibitem{Hacigumus:2002:ESO:564691.564717}
\BIBentryALTinterwordspacing
H.~Hacig\"{u}m\"{u}\c{s}, B.~Iyer, C.~Li, and S.~Mehrotra, ``Executing sql over
  encrypted data in the database-service-provider model,'' in \emph{Proceedings
  of the 2002 ACM SIGMOD international conference on Management of data}, ser.
  SIGMOD '02.\hskip 1em plus 0.5em minus 0.4em\relax New York, NY, USA: ACM,
  2002, pp. 216--227. [Online]. Available:
  \url{http://doi.acm.org/10.1145/564691.564717}
\BIBentrySTDinterwordspacing

\bibitem{antonopoulos2020azure}
P.~Antonopoulos, A.~Arasu, K.~D. Singh, K.~Eguro, N.~Gupta, R.~Jain,
  R.~Kaushik, H.~Kodavalla, D.~Kossmann, N.~Ogg \emph{et~al.}, ``Azure sql
  database always encrypted,'' in \emph{Proceedings of the 2020 ACM SIGMOD
  International Conference on Management of Data}, 2020, pp. 1511--1525.

\bibitem{jagadeeswaraiah2015securedbaas}
P.~Jagadeeswaraiah and M.~P. Kumar, ``Securedbaas model for accessing encrypted
  cloud databases,'' \emph{Telkomnika Indonesian Journal of Electrical
  Engineering}, vol.~16, no.~2, pp. 333--340, 2015.

\bibitem{AWS_data_at_rest}
\BIBentryALTinterwordspacing
K.~Beer and R.~Holland, ``Securing data at rest with encryption.'' [Online].
  Available:
  \url{https://d0.awsstatic.com/whitepapers/aws-securing-data-at-rest-with-encryption.pdf}
\BIBentrySTDinterwordspacing

\bibitem{sym_token}
\BIBentryALTinterwordspacing
Symantec, ``Data privacy and compliance in the cloud.'' [Online]. Available:
  \url{https://www.symantec.com/content/dam/symantec/docs/white-papers/data-privacy-and-compliance-in-the-cloud-en.pdf}
\BIBentrySTDinterwordspacing

\bibitem{schikuta1998vipios}
E.~Schikuta, T.~Fuerle, and H.~Wanek, ``Vipios: The vienna parallel
  input/output system,'' in \emph{Euro-Par98 Parallel Processing: 4th
  International Euro-Par Conference}.\hskip 1em plus 0.5em minus 0.4em\relax
  Springer, 1998, pp. 953--958.

\bibitem{mach2012generic}
W.~Mach and E.~Schikuta, ``A generic negotiation and re-negotiation framework
  for consumer-provider contracting of web services,'' in \emph{Proceedings of
  the 14th International Conference on Information Integration and Web-based
  Applications \& Services}, 2012, pp. 348--351.

\bibitem{schikuta2004n2grid}
E.~Schikuta and T.~Weish{\"a}upl, ``N2grid: neural networks in the grid,'' in
  \emph{2004 IEEE International Joint Conference on Neural Networks (IEEE Cat.
  No. 04CH37541)}, vol.~2.\hskip 1em plus 0.5em minus 0.4em\relax IEEE, 2004,
  pp. 1409--1414.

\bibitem{cs745}
T.~Weish{\"a}upl, F.~Donno, E.~Schikuta, H.~Stockinger, and H.~Wanek,
  ``Business in the grid: The big project,'' in \emph{Grid Economics \&
  Business Models (GECON 2005) of Global Grid Forum}, vol.~13, 2005.

\bibitem{weishaeupl2004}
T.~Weish{\"a}upl and E.~Schikuta, ``Towards the merger of grid and economy,''
  in \emph{Grid and Cooperative Computing - GCC 2004 Workshops}, H.~Jin,
  Y.~Pan, N.~Xiao, and J.~Sun, Eds.\hskip 1em plus 0.5em minus 0.4em\relax
  Berlin, Heidelberg: Springer Berlin Heidelberg, 2004, pp. 563--570.

\bibitem{schikuta2008grid}
E.~Schikuta, H.~Wanek, and I.~Ul~Haq, ``Grid workflow optimization regarding
  dynamically changing resources and conditions,'' \emph{Concurrency and
  Computation: Practice and Experience}, vol.~20, no.~15, pp. 1837--1849, 2008.

\bibitem{kofler2009parallel}
K.~Kofler, I.~Ul~Haq, and E.~Schikuta, ``A parallel branch and bound algorithm
  for workflow qos optimization,'' in \emph{2009 International Conference on
  Parallel Processing}.\hskip 1em plus 0.5em minus 0.4em\relax IEEE, 2009, pp.
  478--485.
  
\newpage


\bibitem{stuermer2009building}
G.~Stuermer, J.~Mangler, and E.~Schikuta, ``Building a modular service oriented
  workflow engine,'' in \emph{2009 IEEE international conference on
  service-oriented computing and applications (SOCA)}.\hskip 1em plus 0.5em
  minus 0.4em\relax IEEE, 2009, pp. 1--4.

\bibitem{citeulike:9349465}
\BIBentryALTinterwordspacing
A.~Boldyreva, N.~Chenette, Y.~Lee, and A.~O'Neill, ``{Order-Preserving
  Symmetric Encryption},'' in \emph{Proceedings of the 28th Annual
  International Conference on Advances in Cryptology: the Theory and
  Applications of Cryptographic Techniques}, ser. EUROCRYPT '09.\hskip 1em plus
  0.5em minus 0.4em\relax Berlin, Heidelberg: Springer-Verlag, 2009, pp.
  224--241. [Online]. Available:
  \url{http://dx.doi.org/10.1007/978-3-642-01001-9\_13}
\BIBentrySTDinterwordspacing

\bibitem{Dworkin2013}
D.~Morris, ``Recommendation for block cipher modes of operation: methods for
  formatpreserving encryption,'' \emph{NIST Special Publication}, vol. 800, p.
  38G, 2013.

\bibitem{heron2009advanced}
S.~Heron, ``Advanced encryption standard (aes),'' \emph{Network Security}, vol.
  2009, no.~12, pp. 8--12, 2009.

\bibitem{hoofnagle2019european}
C.~J. Hoofnagle, B.~Van Der~Sloot, and F.~Z. Borgesius, ``The european union
  general data protection regulation: what it is and what it means,''
  \emph{Information \& Communications Technology Law}, vol.~28, no.~1, pp.
  65--98, 2019.

\end{thebibliography}


\end{document}